\documentclass[twocolumn,secnumarabic, amsmath,amssymb, aps, pre]{revtex4-2}
\usepackage{graphicx}
\usepackage{bm}
\usepackage{multirow}
\usepackage{color}
\newcommand{\pN}[0]{\rm{p N}}\newcommand{\m}[0]{\rm{m}}\newcommand{\unitfric}[0]{\pN \sec / \mu\m}\begin{document}\title{Pattern formation and the mechanics of a motor-driven filamentous system confined by rigid membranes }\author{Mitsusuke Tarama}\email[]{tarama.mitsusuke@phys.kyushu-u.ac.jp}\affiliation{Department of Physics, Kyushu University, Fukuoka 819-0395, Japan}\affiliation{Laboratory for Physical Biology, RIKEN Center for Biosystems Dynamics Research, Kobe 650-0047, Japan}\author{Tatsuo Shibata}\affiliation{Laboratory for Physical Biology, RIKEN Center for Biosystems Dynamics Research, Kobe 650-0047, Japan}\date{\today}\begin{abstract}Pattern formation and the mechanics of a mixture of actin filaments and myosin motors that is confined by a rigid membrane is investigated. By using a coarse-grained molecular dynamics model, we demonstrate that the competition between the depletion force and the active force of the motors gives rise to actin accumulation in the membrane vicinity. The resulting actomyosin structure exerts pressure on the membrane, that, due to nematic alignment of the filaments, converges to a constant for large motor active force. The results are independent of filament length and membrane curvature, indicating the universality of this phenomenon. Thus, this study proposes a novel mechanism by which the compounds of the cytoskeleton can self-organize into a higher-order structure. \end{abstract}\maketitle\section{Introduction} \label{sec:introduction}The field of active matter has seen significant development in recent decades. Seminal studies include the finding of the ordered state in the Vicsek model \cite{Vicsek1995Novel}, the theoretical study on Purcel's swimmer \cite{Najafi2004Simple}, and the experimental realization of active colloids \cite{Paxton2004Paxton,Paxton2006Chemical,Ebbens2010In}. Following these, a number of studies have found that the active system can self-organize into various patterns \cite{Vicsek2012Collective,Ramaswamy2010The}. In addition, many studies, such as the ones on suspensions of microorganisms \cite{Sokolov2009Reduction,Gachelin2013Non-Newtonian}, examined the rheological properties, that are affected by coordinated activity. These studies suggest that the pattern formation observed in active matter systems may build a basis for complex emergent functions \cite{Gompper2020The}. \par In fact, the mechanical functions of self-organized structures are of fundamental importance in biological systems. For instance, cells that exhibit spontaneous motion such as migration and shape deformation \cite{Ohta2016Simple,Tarama2018Mechanics,Tarama2019Mechanochemical} coordinate their motion to close wounds and to enable tissue morphogenesis \cite{Alert2021Living,Saw2017Topological,Howard2011Turings,Takeda2018A}. In these processes, the major force-generating component in cells is cytoskeleton, such as actin filaments and microtubules \cite{Small1998Assembling, Paluch2006Cracking, Pullarkat2007Rheological, Lecuit2011Force, Huber2013Emergent, Banerjee2020The}. These are considered composite active matter as they exhibit active dynamics through the force generated by their associated motor proteins. They form a rich variety of structures, such as the actomyosin cortex \cite{Chugh2018The, Salbreux2012Actin}, asters \cite{Smith2007Molecular,Koster2016Actomyosin,Fritzsche2017Self-organizing}, clusters \cite{Koenderink2009An}, vortices \cite{Nedelec1997Self-organization,Surrey2001Physical,Kruse2004Asters}, stress fibers \cite{Hotulainen2006Stress, Burridge2016Focal, Hu2017Long-range, Peterson2004Simultaneous, Tojkander2012Actin}, contractile rings \cite{Pelham2002Actin, Barr2007Cytokinesis}, polarity sorting \cite{Sciortino2021Pattern}, and mitotic spindles \cite{Bennabi2016Meiotic}, which possess characteristic functions \cite{Trepat2007Universal, Sackmann2010Physics, Shah2013Mechanical, Blanchoin2014Actin, Heer2017Tension, Svitkina2018Ultrastructure, Carlsson2018Membrane}. \par Among them, the actomyosin cortex is one of the most basic structures, as it maintains the cellular shape. It is a network underneath the cellular membrane, consisting of actin filaments and myosin motors as well as other associated proteins. Thus, to form the actomyosin cortex, these molecules first need to accumulate in the membrane vicinity. Interestingly, such accumulation of active matter under confinement has been reported in recent studies using very simple models of active colloids \cite{Smallenburg2015Swim, Elgeti2013Wall} and active rods \cite{Abaurrea-Velasco2017Complex,Abaurrea-Velasco2019Vesicles}. The phenomenon has been compared with motility-induced phase separation \cite{Cates2015Motility-Induced, Bechinger2016Active}. However, while actomyosin and active colloids show similar active dynamics, there are significant differences between these systems. Firstly, unlike active colloids, actin filaments do not self-propel unless they treadmill (an effect not considered in our study) \cite{note_treadmill}. Secondly, actin filaments are very long compared to the typical active colloidal rods. They are filamentous molecules of about 10 nm in width and a few tens of nm to several $\mu \m$ in length \cite{Mueller2017Load}. In addition, their persistence length is about 10 $\mu\m$, comparable to the size of a cell, which also makes them quite stiff. Generally, such a long stiff object experiences a strong depletion force when confined due to the resulting restriction of its rotational degree of freedom. Therefore, in the case of these long stiff objects that only becomes active when bound by the motors that produce force on them, accumulation at a confining membrane is both non-trivial yet highly relevant for biological systems. \par The purpose of this study is to investigate pattern formation of an actomyosin network that interacts with a confining membrane and to analyze the mechanical function of the resulting self-organized structure. In particular, we focus on the competition between the depletion force and the motor active force. Previously, the dynamics of actomyosin has been studied by using a macroscopic continuum model referred to as active gel model \cite{Prost2015Active, Kruse2006Contractility, Julicher2007Active}. However, in order to bridge the microscopic molecular information and macroscopic structures, we develop a mesoscopic molecular dynamics model, in a similar spirit of those in Refs~\cite{Astrom2008Strain,Ziebert2008Rheological}, in which the molecular origin of the macroscopic mechanics were analyzed. \par This paper is organised as follows. We first define the coarse-grained molecular dynamics model of cytoskeleton in the next section, which is followed by the explanation of the simulation results in section~\ref{sec:results}. Finally, Sec.~\ref{sec:discussion} is devoted to the summary and discussion. \section{Model} \label{sec:model}\begin{figure}[b]\centering\includegraphics{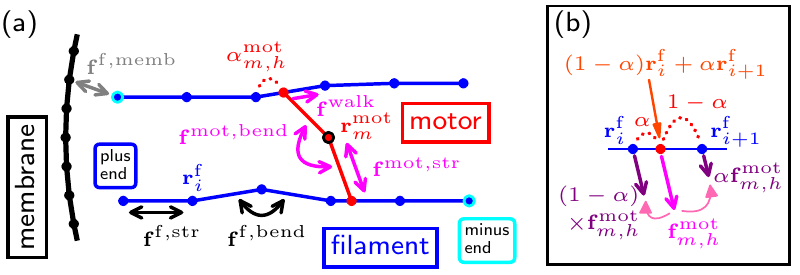}\caption{(a) Schematics of the coarse-grained model of filaments and motors confined by a membrane. (b) Sketch of the interpolation between the filament particles, on which the motor head force is assigned using the coordinate $\alpha$. }\label{fig:model}\end{figure}\par We start by defining our coarse-grained molecular dynamics model of actin filaments and myosin motors.  We model a filament by a linear series of discrete particles connected by elastic springs and a motor by a particle with two heads (Fig.~\ref{fig:model}). The two motor heads can bind to two different filaments and walk actively along them. This reproduces the motion of a bipolar motor such as the non-muscle myosin II minifilaments that possess multiple actin binding sites on both ends \cite{Vasquez2016Drosophila}. The equations of motion for the filament and motor particles are given by \begin{gather}\gamma^{\rm f} \frac{d \bm{r}^{\rm f}_{f,i}}{dt} = \bm{f}^{\rm f,str}_{f,i} +\bm{f}^{\rm f,bend}_{f,i} +\bm{f}^{\rm f,mot}_{f,i} +\bm{f}^{\rm f,memb}_{f,i} +\bm{\xi}^{\rm f}_{f,i} \label{eq:drdt:filament} \\\gamma^{\rm mot} \frac{d \bm{r}^{\rm mot}_m}{dt} = \sum_{h=0,1} \big( -\bm{f}^{\rm mot,str}_{m,h} -\bm{f}^{\rm mot,bend}_{m,h} \big) +\bm{\xi}^{\rm mot}_m \label{eq:drdt:motor_body}\end{gather}Here $\bm{r}^{\rm f}_{f,i}$ represents the position of the $i$th particle of filament $f$. The index $f$ is omitted hereafter for ease of notation. $\bm{r}^{\rm mot}_m$ is the position of the motor $m$, and the coordinate $0 \le \alpha^{\rm mot}_{m,h} \le 1$ along the filament segment $\bm{r}^{\rm f}_i$--$\bm{r}^{\rm f}_{i+1}$ gives the position of the $h$th head of motor $m$ as $(1-\alpha^{\rm mot}_{m,h}) \bm{r}^{\rm f}_{i} +\alpha^{\rm mot}_{m,h} \bm{r}^{\rm f}_{i+1}$ (Fig.~\ref{fig:model}). Then, the equation of motion for the motor head reads \begin{gather}\zeta r^{\rm f}_{i(i+1)} \frac{d \alpha^{\rm mot}_{m,h}}{dt} = f^{\rm walk},  \label{eq:drdt:motor_head}\end{gather}where $r^{\rm f}_{i(i+1)} = | \bm{r}^{\rm f}_{i+1} -\bm{r}^{\rm f}_{i} |$ is the filament segment length. See Appendix~\ref{sec:EOM_motor} for the derivation. Inertia terms are neglected because of the small size and velocity of the molecules; The typical sizes of the cytoskeletal filament and the motor are less than a few $\mu \m$ and submicrometers, respectively. The sliding velocity of the motor head is submicrometers per second. Note that the filament polarity is taken into account by the order of the filament particle indices, which gives the cue for the motor heads to walk actively towards the filament plus end, i.e., in increasing index $i$. The filament and motor particles experience friction and fluctuation from the surrounding cytosolic fluid, which satisfy the fluctuation-dissipation theorem; $\langle \bm{\xi}^{\rm X}_i \rangle = 0$ and $\langle \xi^{\rm X}_{i,a}(0) \xi^{\rm X}_{j,b}(t) \rangle = 2 \gamma^{\rm X} \delta_{ij} \delta_{ab} \delta(t)$ with ${\rm X} = \{{\rm f},{\rm mot}\}$. The cytosolic friction is defined using a cylindrical approximation \cite{Kim2009Computational} as $\gamma^{\rm f} =  3\pi \eta_{\rm cyto} (3 d^{\rm f} +2 \ell^{\rm f}_0/5)$ and $\gamma^{\rm mot} =  6\pi \eta_{\rm cyto} (3 d^{\rm mot} +\ell^{\rm mot}_0 /5)$, where $\eta_{\rm cyto}$ is the cytosolic viscousity and $d^{\rm f}$ ($d^{\rm mot}$) and $\ell^{\rm f}_0$ ($\ell^{\rm mot}_0$) are the width and length of the filament segment (motor), respectively. On the other hand, since the motor heads are bound to and walk along filaments due to the motor walk force (MWF) $f^{\rm walk}$, they experience a sliding friction $\zeta r^{\rm f}_{i(i+1)} d \alpha^{\rm mot}_{m,h}/dt$. \par The immobile membrane is also discretized by small particles \cite{Tarama2017Swinging}, which confine the filament through the repulsive interaction\begin{align}f^{\rm f,memb}_{i} = -\frac{\partial }{\partial \bm{r}^{\rm f}_{i}} U^{\rm rep} ( \left| \bm{r}^{\rm memb}_k -\bm{r}^{\rm f}_{i} \right|; \epsilon^{\rm rep}, \sigma^{\rm rep}, r^{\rm rep}_{\rm *} ) \label{eq:f^rep_memb} \end{align} with $U^{\rm rep} ( r; \epsilon, \sigma, r_{*} ) = \frac{\epsilon}{1- \sigma /r_{*}} \left( \sigma/r -\sigma/r_{*} \right) H( r_{*} -r )$. $\epsilon^{\rm rep}$, $\sigma^{\rm rep}$, and $r^{\rm rep}_{*}$ are the intensity, decay length, and cutoff length of the repulsive interaction. $\bm{r}^{\rm memb}_k$ is the position of the membrane particle closest to the filament particle $\bm{r}^{\rm f}_{i}$. $H(x)$ is the Heaviside step function that takes the value 1 for $x>0$ and 0 otherwise. \par The length and straightness of the filament are maintained by the stretching and bending elasticity acting between the filament particles: \begin{gather}\bm{f}^{\rm f,str}_{i} = - \frac{\partial}{\partial \bm{r}^{\rm f}_{i}}\sum_i U^{\rm harm} ( r^{\rm f}_{i(i+1)}; \kappa^{\rm f,str}, \ell^{\rm f}_0 ),\label{eq:Ustr_filament_harmonic}\\ \bm{f}^{\rm f,bend}_{i} = - \frac{\partial}{\partial \bm{r}^{\rm f}_{i}} \sum_i U^{\rm bend}( \bm{r}^{\rm f}_{(i-1)i}, \bm{r}^{\rm f}_{i(i+1)}; k^{\rm f,bend} ), \label{eq:U^bend_filament}\end{gather}where $U^{\rm harm} ( r ; \kappa, \ell_0 )$ is a harmonic potential with elastic modulus $\kappa$ and rest length $\ell_0$, and $U^{\rm bend} ( \bm{r}_{-}, \bm{r}_{+} ; k ) = - k \hat{\bm{r}}_{-} \cdot \hat{\bm{r}}_{+}$ is bending energy with bending rigidity $k^{\rm f,bend}$. Here, we use the abbreviations $r = |\bm{r}|$, $\bm{r}^{\rm f}_{ij} = \bm{r}^{\rm f}_{j} -\bm{r}^{\rm f}_{i}$, and $\hat{\bm{x}} = \bm{x} / | \bm{x} |$. \par Similarly, the stretching and bending forces act between the motor particles and motor heads: \begin{gather}	\bm{f}^{\rm mot,str}_{m,h} = - \frac{\partial}{\partial \bm{r}^{\rm mot}_{m,h}} \sum_{h} U^{\rm harm}( l^{\rm mot}_{m,h}; \kappa^{\rm mot,str}, \ell^{\rm mot}_0 /2 )\label{eq:f_motor_stretching}\\ \bm{f}^{\rm mot,bend}_{m,h} = -\frac{\partial}{\partial \bm{r}^{\rm mot}_{m,h}} U^{\rm bend}( -\bm{l}^{\rm mot}_{m,0}, \bm{l}^{\rm mot}_{m,1}; k^{\rm mot,bend} ), \label{eq:f^bend_motorhead}\end{gather}where $\bm{l}^{\rm mot}_{m,h} = \bm{r}^{\rm mot}_{m,h} -\bm{r}^{\rm mot}_{m}$. Their contribution to the motor heads, $\bm{f}^{\rm mot}_{m,h} = \bm{f}^{\rm mot,str}_{m,h}  +\bm{f}^{\rm mot,bend}_{m,h}$, is assigned to the filament particles with the geometric weight $\alpha^{\rm mot}_{m,h}$, which is necessary because the binding position may be in between the filament particles (Fig.~\ref{fig:model}b). Therefore, the force on the filament particle $i$ from the motors reads $\bm{f}^{\rm f,mot}_i = \sum_{m,h} ( 1 -\alpha^{\rm mot}_{m,h} ) \bm{f}^{\rm mot}_{m,h} - \sum_{m,h} \alpha^{\rm mot}_{m,h}  \bm{f}^{\rm mot}_{m,h}$. Here the first and second summations are calculated over the motor heads bound to the filament segments $\bm{r}^{\rm f}_i$--$\bm{r}^{\rm f}_{i+1}$ and $\bm{r}^{\rm f}_{i-1}$--$\bm{r}^{\rm f}_{i}$, respectively. This ensures force and torque conservation of the filament segments. Note that the counter force of the motor sliding friction and MWF, which act on the filaments, vanishes because of the force balance equation \eqref{eq:drdt:motor_head}. Therefore, Eqs.~\eqref{eq:drdt:filament}--\eqref{eq:drdt:motor_head} statistically satisfy the force- and torque-free conditions, which are required for active systems including migrating cells~\cite{Tarama2018Mechanics,Tarama2019Mechanochemical}. \par In addition to the equations of motion \eqref{eq:drdt:filament}-\eqref{eq:drdt:motor_head}, we consider the following stochastic processes. The first one is the process of motor binding and unbinding. Unbinding of a motor head occurs in three cases. Firstly, a motor head that is bound to a filament unbinds stochastically at the rate of $\omega^{\rm mot}_{\rm TO}$. Secondly, when a motor head reaches the end of a filament by actively walking along it, it unbinds instantaneously with the probability unity. Thirdly, a motor head unbinds with the probability unity when the other head of the motor unbinds. Here we assume that a motor takes either the bound state where both of the heads are bound to filaments or the free state where neither of the two heads are bound to filaments for simplicity. The free motors are assumed to diffuse sufficiently fast, so that they distribute uniformly. Then, a free motor binds to a randomly-selected pair of filaments whenever they find a position on each filament that are separated by the length of the motor $\ell^{\rm mot}_0$. Thus, this motor binding process alone causes no force on the system. The active force is generated in the cytoskeleton through the motor stretching and bending energies that are stored when motor heads move along filaments. The second stochastic process is filament turnover that takes place at a rate of $\omega^{\rm f}_{\rm TO}$. The filament undergoing turnover is placed back into the system immediately at random position with random orientation, and all the motors previously bound to it become free. Although this model can be applied to both actin filaments and microtubules, in this paper, we focus on the system of actin filaments and myosin motors with an appropriate choice of the relevant parameters as summarized in Appendix~\ref{sec:parameters}. \par We solve the set of time-evolution equations in the following manner; First, we calculate the turnover of the filaments and motors, and the position of the motor head is updated by solving Eq.~\eqref{eq:drdt:motor_head} with the Euler method. Then, Eqs.~\eqref{eq:drdt:filament} and \eqref{eq:drdt:motor_body} are solved by using the fourth-order Runge-Kutta method. In the following, energy, length, and time are rescaled by using thermal energy $U_0=k_B T$, motor length $l_0=\ell^{\rm mot}_0$, and motor cytosolic friction coefficient $t_0 = \gamma^{\rm mot} (\ell^{\rm mot}_0)^2 /k_B T$. \par Finally we comment on the difference of this approach from existing models. There are several open source packages based on similar molecular dynamics models, including AFINES \cite{Freedman2017A}, aLENS \cite{Yan2021aLENS}, CyLaKS \cite{Fiorenza2021CyLaKS}, Cytosim \cite{Nedelec2007Collective}, and Medyan \cite{Popov2016MEDYAN}. In these models, a motor is reduced to a harmonic potential with zero rest length or it is represented as a single segment with finite length since myosin motors often form minifilaments. In constrast to these pre-packaged simulation approaches, a motor in our model is modeled by two segments, which connect the motor particle with the two heads. This treatment is very similar to those in Ref.~\cite{Kim2009Computational,Matsuda2019Mobility} and allows to include both the finite length of the motor and its cytosolic friction, which is mechanically more relevant to the actual situation since the cytosolic friction depends on the length of the motor. \section{Results} \label{sec:results}\subsection{Accumulation in the membrane vicinity}\label{sec:accumulation}\begin{figure*}[t]\centering\includegraphics{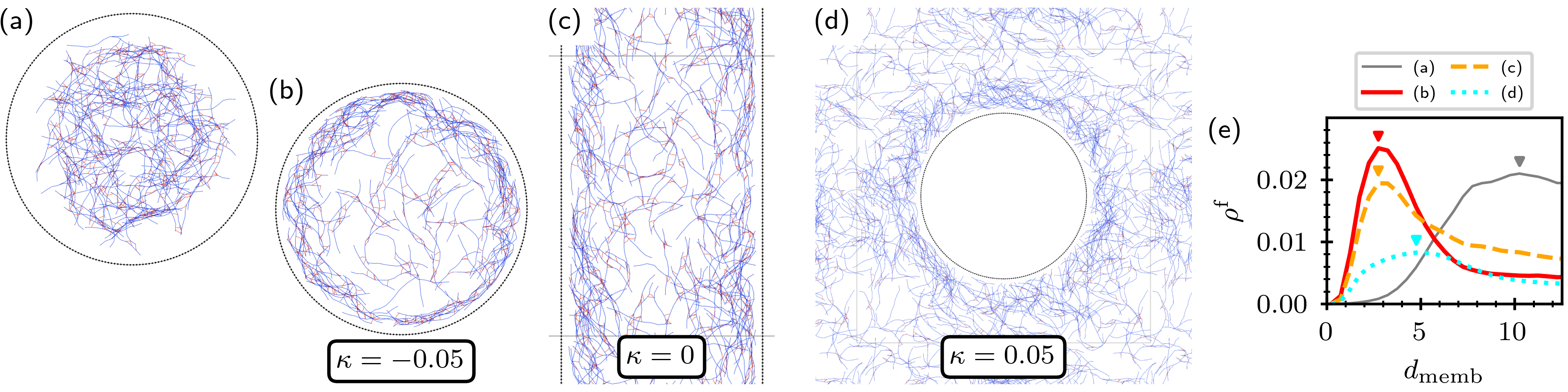}\caption{Snapshots of the filaments (blue) and motors (red) confined by the membrane (black). (a) A depletion zone of the filaments of length $L_{\rm f}=10$ is observed in the absence of motors. (b--d) Cortex-like structure is formed around the membrane with curvature (b) $\kappa=-0.05$, (c) 0, and (d) $0.05$. The number of filaments and motors are (a--c) $(N_{\rm f},N_{\rm m})=(200,600)$ and (d) $(600,600)$. The motor walk force (MWF) is set to $f^{\rm walk} = 0$ in (a) and $\approx 2.4$ in (b--d), which gives the estimated speed of $0.5 \mu\m/\sec$ consistent with experimental measurements. In panels (c,d), thin gray lines indicate the periodic boundaries. (e) Density distribution of filaments (DDF) $\rho^{\rm f}$ as a function of the distance from the membrane $d_{\rm memb}$ with its first peak position indicated by arrow heads. The cutoff distance of the filament-membrane repulsive interaction is set to $r_{*} = 1.05$. The lines correspond to the snapshots in (a--d). }\label{fig:snapshot}\end{figure*}\par First, we consider actomyosin dynamics inside a circular membrane. Without motors, the filaments are depleted near the membrane. The depletion zone is also observed when motors are introduced but no MWF is applied (Fig.~\ref{fig:snapshot}a). For finite MWF, however, the filaments accumulate in the vicinity of the membrane, forming a structure that resembles the actomyosin cortex (Fig.~\ref{fig:snapshot}b). Here, the membrane curvature is set to $\kappa=0.05$, corresponding to a radius $R=20$ ($4 \mu\m$), which is twice the filament length $L_{\rm f}=10$ ($2 \mu\m$). \par To investigate whether the observed accumulation is universal or a simple consequence of the confinement curvature, we varied the curvature of the membrane. Interestingly, the filament accumulation is found for planar membranes with zero curvature and even for an oppositely curved circular membrane with positive curvature, in which case the filaments are present outside the membrane (Figs.~\ref{fig:snapshot}c and \ref{fig:snapshot}d). \begin{figure*}[tbh]\centering\includegraphics{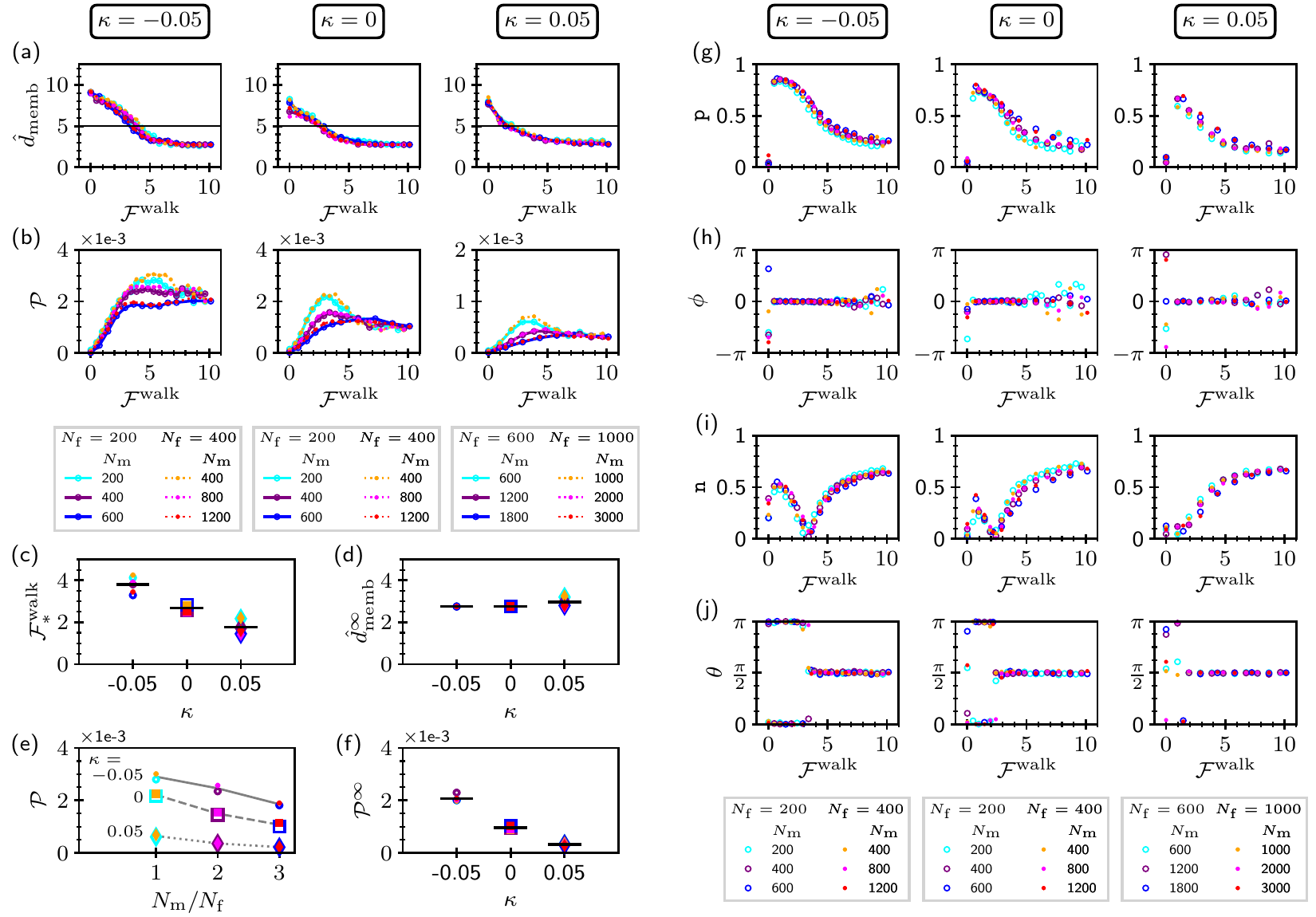}\caption{Cortex-like structure formation and force generation due to MWF. (a) DDF peak distance $\hat{d}_{\rm memb}$ shows a collapse of data as a function of the average MWF $\mathcal{F}^{\rm walk}$. The black horizontal lines indicate $L_{\rm f} /2$. (b) Scaled pressure $\mathcal{P}$ against $\mathcal{F}^{\rm walk}$. Dependence of (c) $\mathcal{F}^{\rm walk}_*$ and (d) $\hat{d}_{\rm memb}^{\infty} $ on the membrane curvature $\kappa$. (e) Scaled pressure $\mathcal{P}$ against $N_{\rm m}/N_{\rm f}$ for the intermediate $\mathcal{F}^{\rm walk}$ ($\approx 5.8$ for $\kappa=-0.05$ and $2.9$ for $\kappa=0$ and $0.05$). (f) Dependence of $\mathcal{P}^{\infty}$ on the membrane curvature $\kappa$. In panels (c,d,f), the average values are indicated by the black horizontal bars for each $\kappa$. The circle, square, and diamond symbols in panels (c-f) represent the data for $\kappa=-0.05$, 0, and 0.05, and the color of the symbols distinguishes different $N_{\rm f}$ and $N_{\rm m}$ corresponding to those in panels (a-b). (g--j) Structural order parameters of filaments in the membrane vicinity. (g) Magnitude $\mathtt{p}$ and (h) direction $\phi$ of the polar order. (i) Magnitude $\mathtt{n}$ and (j) angle $\theta$ of the nematic order. The columns in panels (a,b,g-j) correspond to the membrane curvatures $\kappa=-0.05$ (left), 0 (middle), and $0.05$ (right), respectively. }\label{fig:structure}\end{figure*}\par To quantify this accumulation, we calculated the density distribution of filaments (DDF) $\rho^{\rm f}$ defined as a function of the distance from the membrane $d_{\rm memb}$ by\begin{align}\rho^{\rm f}(d_{\rm memb}) = \Big\langle\frac{1}{A(d_{\rm memb})} \sum_{i} \delta_{i}(d_{\rm memb}; d'_d) \Big\rangle_t,\label{eq:rho^f}\end{align}where the function $\delta_{f,i}(d_{\rm memb}; d'_d)$ is 1 if the filament segment $\bm{r}^{\rm f}_i$--$\bm{r}^{\rm f}_{i+1}$ is at the distance $[d_{\rm memb} -d'_d/2, d_{\rm memb} +d'_d/2 ]$ from the membrane and equals 0 otherwise: \begin{align}&\delta_{i}(d_{\rm memb})\notag \\ &= 1 - H( d_{\rm memb} -\frac{d'_d}{2} -d^{\rm f}_{i} ) H( d^{\rm f}_{i} -d_{\rm memb} -\frac{d'_d}{2} )\notag\\&\times H( d_{\rm memb} -\frac{d'_d}{2} -d^{\rm f}_{i+1} ) H( d^{\rm f}_{i+1} -d_{\rm memb} -\frac{d'_d}{2} ).\label{eq:rho^f_detail}\end{align}$A(d_{\rm memb})$ is the area of the region within the distance $d_{\rm memb} -d'_d/2 \le r \le d_{\rm memb} +d'_d/2$ from the membrane, and $d'_d$ gives the width of the average region, which we set to $d'_d = \ell^{\rm f}_0 /2$. $\langle x \rangle_t$ represents the time average. Since the filaments can bend, we calculate the DDF using the distance of all filament particles from the membrane $d^{\rm f}_{i} = ( \bm{r}^{\rm f}_{i} -\bm{r}^{\rm memb}_k ) \cdot \bm{n}_k$, where $\bm{n}_k$ is the normal direction to the membrane. \par The DDF reaches zero at the membrane and increases with $d_{\rm memb}$ (Fig.~\ref{fig:snapshot}e). When MWF is absent, $\rho^{\rm f}$ almost vanishes at small $d_{\rm memb}$ showing a peak at larger distance, which represents the fact that the filaments are depleted from the membrane vicinity. When MWF is switched on, however, the peak in $\rho^{\rm f}$ appears closer to the membrane, as marked by the arrow heads. This indicates the filament accumulation in the membrane vicinity. \par At equilibrium, filaments are depleted from the membrane vicinity where their rotational degrees of freedom are restricted (See Appendix~\ref{sec:depletion}). This is also true when motors cross-link filaments but lack the active force generation (Fig.~\ref{fig:snapshot}a). In order for filaments to accumulate in the membrane vicinity, MWF needs to overcome the depletion force. To quantify this point, we plotted the DDF peak distance, $\hat{d}_{\rm memb}$, as a function of MWF. Although $\hat{d}_{\rm memb}$ also depends on the number of filaments and motors, we found that all data collapse on top of each other when the MWF is rescaled to the average MWF $\mathcal{F}^{\rm walk} = f^{\rm walk} N_{\rm m} / N_{\rm f}$ (Fig.~\ref{fig:structure}a). This indicates the universality of the observed phenomena. As $\mathcal{F}^{\rm walk}$ increases, $\hat{d}_{\rm memb}$ shifts towards the membrane and becomes smaller than $L^{\rm f}/2$, at which the depletion force sets in, and thus, $\hat{d}_{\rm memb} < L^{\rm f}/2$ corresponds to the filament accumulation in the membrane vicinity. We define the transition point by $\mathcal{F}^{\rm walk} = \mathcal{F}^{\rm walk}_*$ that gives $\hat{d}_{\rm memb} = L^{\rm f}/2$. Note that the critical average MWF is smaller for larger membrane curvature $\kappa$ (Fig.~\ref{fig:structure}c). Presumably this effect occurs because the rotational degree of freedom of the filaments increases with the membrane curvature, which weakens the depletion force (See Fig.~\ref{fig:depletion}c in Appendix~\ref{sec:depletion}). In addition, we note that, $\hat{d}_{\rm memb}$ in the limit of large $\mathcal{F}^{\rm walk}$ converges to a constant value $\hat{d}_{\rm memb}^{\infty}$, which is smaller than $L^{\rm f}/2$ (Fig.~\ref{fig:structure}d). In conclusion, MWF exerted on the cytoskeleton is able to overcome the depletion force, leading to a filament accumulation in the membrane vicinity that resembles the actomyosin cortex. \subsection{Pressure} \label{sec:pressure}One major function of the cytoskeleton is force generation in the cell. Thus, we are interested in the pressure that the actomyosin produces on the membrane as a function of MWF and calculate it by\begin{align}P &= \Big\langle (A_{\rm memb})^{-1} \sum_k \Big( -\frac{\partial U^{\rm rep}}{\partial \bm{r}^{\rm memb}_k} \Big) \cdot \bm{n}_k \Big\rangle_t\label{eq:pressure}\end{align}where $A_{\rm memb}$ is the length (area in 3d) of the membrane. The pressure depends on the number of motors and it scales with the number of filaments $N_{\rm f}$ (Fig.~\ref{fig:structure}b). In fact, all the data of the scaled pressure $\mathcal{P} = P / N_{\rm f}$ collapse except for the intermediate $\mathcal{F}^{\rm walk}$. Peculiarly, in the intermediate regime of $\mathcal{F}^{\rm walk}$, the scaled pressure $\mathcal{P}$ takes a \textit{larger} value for a \textit{smaller} number of motors $N_{\rm m}/N_{\rm f}$, although the data of $\mathcal{P}$ collapses for each ratio $N_{\rm m}/N_{\rm f}$ (Fig.~\ref{fig:structure}e). Another interesting point is that the scaled pressure converges to a constant value $\mathcal{P}^{\infty}$ for large $\mathcal{F}^{\rm walk}$. $\mathcal{P}^{\infty}$ decreases as the membrane curvature increases (Fig.~\ref{fig:structure}f). This convergence means that the active force that the motors generate on the actomyosin accumulation in the membrane vicinity is not directly converted to the pressure on the membrane. \subsection{Filament structure in the accumulation} \label{sec:structure}To understand the reason why the pressure on the membrane become a constant for large MWF, we investigate the structural order of filaments inside the accumulation. To this end, we measure the polar and nematic order of the filaments with respect to the membrane normal direction in the membrane vicinity. The polar and nematic order parameters of the filaments in the membrane vicinity are calculated by\begin{align}\bm{\mathtt{P}} &= \Big\langle \sum_{i} \hat{\bm{r}}^{\rm f}_{i(i+1)} \big(1 -H( d^{\rm f}_{i} -\bar{d}_{\rm memb} ) H( d^{\rm f}_{i+1} -\bar{d}_{\rm memb} ) \big) \Big\rangle_t,\label{eq:polar_vector} \\ \bm{\mathtt{N}} &= \Big\langle \sum_{f,i} \hat{\bm{r}}^{\rm f}_{i(i+1)} \hat{\bm{r}}^{\rm f}_{i(i+1)} \notag\\ &\times\big(1 -H( d^{\rm f}_{i} -\bar{d}_{\rm memb} ) H( d^{\rm f}_{i+1} -\bar{d}_{\rm memb} ) \big) \Big\rangle_t, \label{eq:nematic_tensor}\end{align}where, by considering the relevant neighborhood, the cutoff distance is set to $\bar{d}_{\rm memb} \le 1.25$, which is about half of $\hat{d}_{\rm memb}^{\infty}$. The magnitude and direction of the polar and nematic order parameters are obtained by using the relationship \begin{align}\bm{\mathtt{P}} = \mathtt{p} ( \cos \phi, \sin \phi),~~\bm{\mathtt{N}} = \mathtt{n} \left(\begin{array}{cc}\cos 2\theta & \sin 2\theta \\ \sin 2\theta & -\cos 2\theta	\end{array}\right).\end{align}\par Firstly, the magnitude of the polar order $\mathtt{p}$ decreases as $\mathcal{F}^{\rm walk}$ increases (Fig.~\ref{fig:structure}g), with its direction perpendicular to the membrane $\phi \approx 0$ as shown in Fig.~\ref{fig:structure}h. Secondly, the nematic order $\mathtt{n}$ increases for $\mathcal{F}^{\rm walk} \gtrapprox \mathcal{F}^{\rm walk}_*$ (Fig.~\ref{fig:structure}i), with its direction parallel to the membrane $\theta \approx \pi/2$ as depicted in Fig.~\ref{fig:structure}j. These results indicate that the filaments approach the membrane as a result of MWF for $\mathcal{F}^{\rm walk} \lessapprox \mathcal{F}^{\rm walk}_*$, whereas they tend to align nematically parallel to the membrane for $\mathcal{F}^{\rm walk} \gtrapprox \mathcal{F}^{\rm walk}_*$. This high nematic order of filaments aligning parallel to the membrane is the reason why $\hat{d}_{\rm memb}$ and $\mathcal{P}$ converge to constant values for large $\mathcal{F}^{\rm walk}$ since the parallel filaments can slide along the membrane without pushing it. This also explains why $\mathcal{P}^{\infty}$ decreases as $\kappa$ increases, since for larger membrane curvature the filaments nematically aligned parallel to the membrane have more chance to point away from the membrane (See Fig.~\ref{fig:depletion} in Appendix~\ref{sec:depletion}). Note that the high values of $\mathtt{n}$ for $\mathcal{F}^{\rm walk} \lessapprox \mathcal{F}^{\rm walk}_*$ are induced by the high polar order $\mathtt{p}$, and thus both direction are the same ($\theta \approx \phi \approx 0$). \subsection{Effect of filament length} \label{sec:length}\begin{figure}[tb]\centering\includegraphics{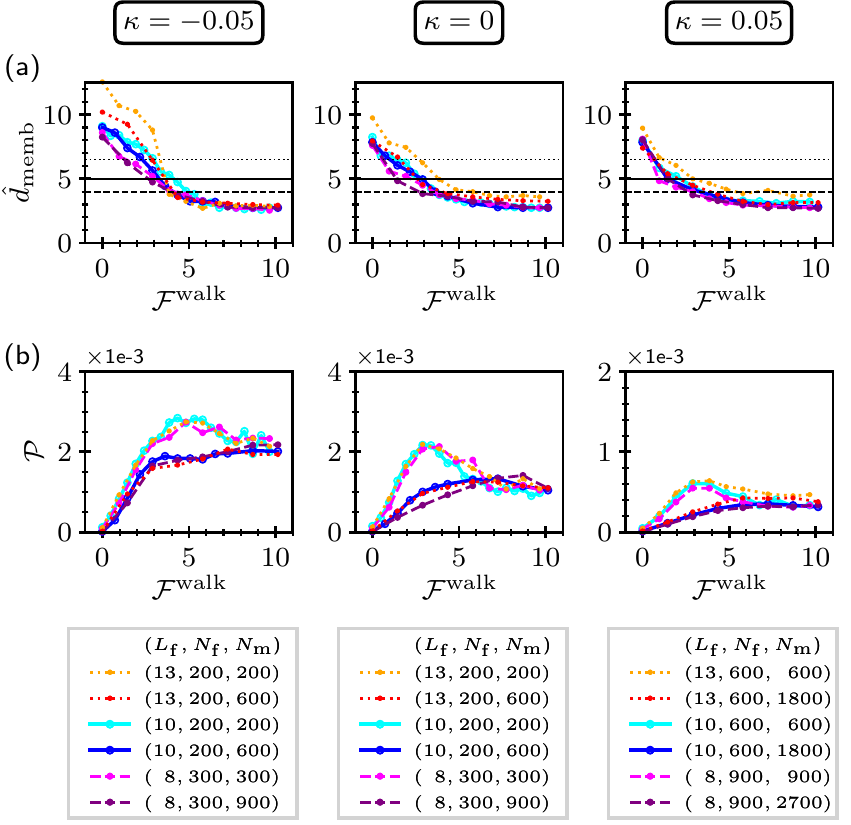}\caption{Dependence of the filament accumulation on the filament length. (a) DDF peak distance $\hat{d}_{\rm memb}$ and (b) scaled pressure on the membrane $\mathcal{P}$ as functions of the average MWF $\mathcal{F}^{\rm walk}$. In panel (a), the dashed, solid, and dotted black lines indicate $L_{\rm f} /2$ for $L_{\rm f} = 8$, 10, and 13, respectively. Columns correspond to the membrane curvatures $\kappa=-0.05$ (left), 0 (middle), and $0.05$ (right), respectively. }\label{fig:filament_length}\end{figure}In this section, we study the impact of the filament length. We performed simulations with shorter ($L_{\rm f}=8$) and longer ($L_{\rm f}=13$) filaments than those used so far ($L_{\rm f}=10$). The results are summarized in Fig.~\ref{fig:filament_length}. The data collapse of both the DDF peak distance $\hat{d}_{\rm memb}$ and the scaled pressure $\mathcal{P}$ is unaltered by the filament length. In particular, the odd increase in $\mathcal{P}$ for smaller motor number at intermediate $\mathcal{F}^{\rm walk}$ is also observed. These results are evidence of the universality of the observed phenomena in the cytoskeleton active dynamics. \begin{figure}[tb]\centering\includegraphics{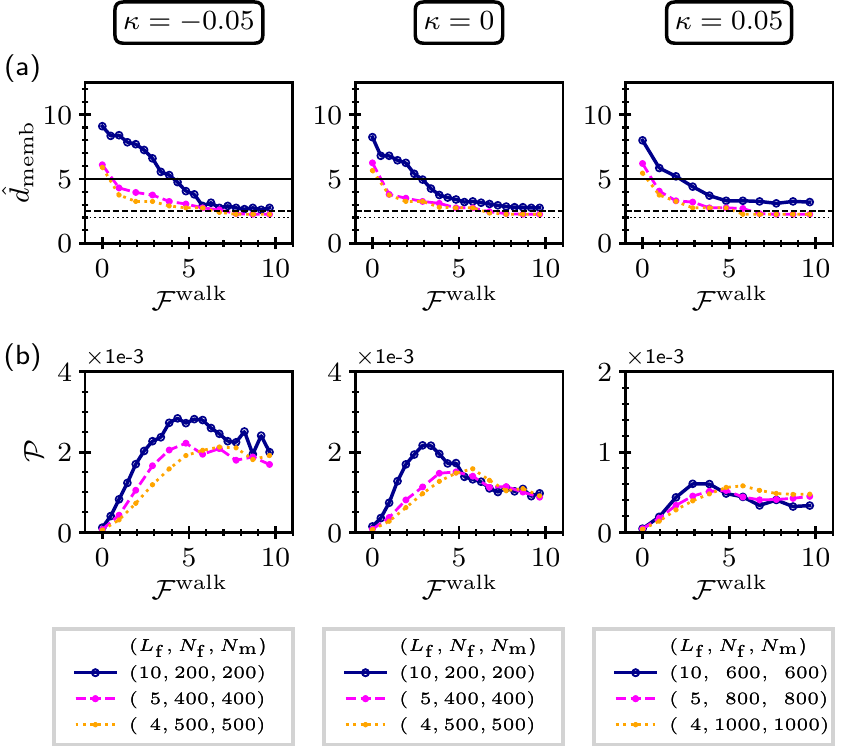}\caption{Filament accumulation of very short filaments. (a) DDF peak distance $\hat{d}_{\rm memb}$ as a function of the average MWF $\mathcal{F}^{\rm walk}$. The solid, dashed, and dotted gray lines indicate $d_{\rm memb} = L_{\rm f} /2$ for $L_{\rm f} = 10$, 5, and 4, respectively. (b) Normalized pressure $\mathcal{P}$. Columns correspond to the membrane curvatures $\kappa=-0.05$ (left), 0 (middle), and $0.05$ (right), respectively. }\label{fig:very_short_filament}\end{figure}\par For even shorter filaments, the collapse of data became worse, although the filament accumulation in the membrane vicinity is still observed for large MWF as shown in Fig.~\ref{fig:very_short_filament}. One possible reason is that the cytosolic friction decreases with filament length, and thus, the cytosolic fluctuation increases. In all cases, however, $\hat{d}_{\rm memb}$ as well as $\mathcal{P}$ still converge to constant values ($\hat{d}_{\rm memb}^{\infty}$ and $\mathcal{P}^{\infty}$) for large $\mathcal{F}^{\rm walk}$. \subsection{Filament motion} \label{sec:motion}\begin{figure*}[t]\centering\includegraphics{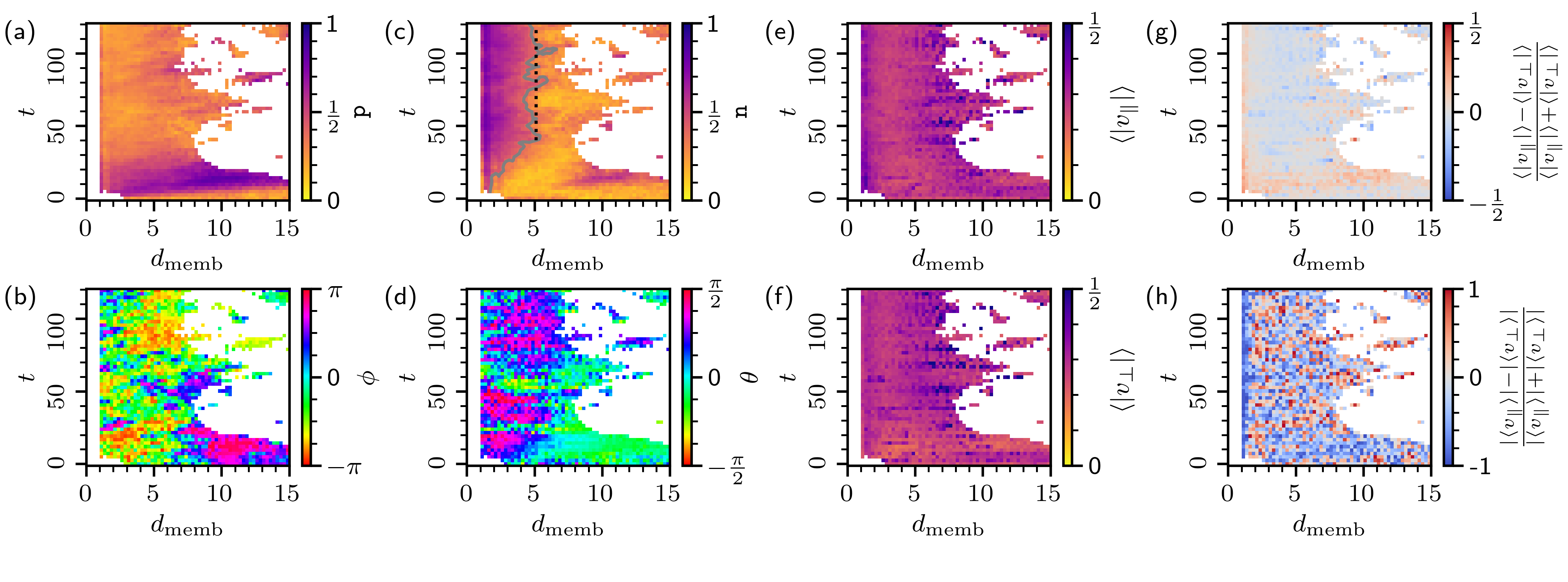}\caption{Time evolution of (a) the polar order parameter $\mathtt{p}$ and (b) its angle $\phi$, (c) the nematic order parameter $\mathtt{n}$ and (d) its angle $\theta$, (e) the tangential $\langle |v_{\parallel}| \rangle$ and (f) normal speed $\langle |v_{\perp}| \rangle$ and (g) their ratio $( \langle |v_{\parallel}| \rangle - \langle |v_{\perp}| \rangle) / ( \langle |v_{\parallel}| \rangle + \langle |v_{\perp}| \rangle|$, and (h) the ratio $( |\langle v_{\parallel} \rangle| - |\langle v_{\perp} \rangle|) / ( |\langle v_{\parallel} \rangle| + |\langle v_{\perp} \rangle|)$ of the tangential $\langle v_{\parallel} \rangle$ and normal velocity $\langle v_{\perp} \rangle$. In panel (c), the boundary of the high nematic region is displayed by the gray solid line, and its time-average at the steady state ($t \ge 40$) is displayed by the black dotted line. The data corresponds to the case plotted in Fig.~\ref{fig:snapshot}b. The kymographs are displayed in the region where $\rho^{\rm f} > 0.01$. }\label{fig:density_time_evolution}\end{figure*}Finally, to understand how the filaments move and accumulate in the membrane vicinity, we plot the time evolution of the polar and nematic order parameters, and the ratio of the tangential and normal speed and velocity (i.e., signed speed) with respect to the membrane in Fig.~\ref{fig:density_time_evolution}. \par Starting from a uniform distribution, filaments first form polar order in the bulk, i.e., far from the membrane (purple region around $t \lesssim 20$ in Fig.~\ref{fig:density_time_evolution}a). This drives the filaments towards the membrane with a velocity given by the perpendicular component of the speed $\langle |v_{\perp}| \rangle$. At the same time, the filaments also move in the parallel direction to the membrane as indicated by the high value of the tangential speed $\langle |v_{\parallel}| \rangle$. In fact, these two components are comparable in strength (with the tangential component slightly higher than the normal one), causing a diagonal motion. In contrast, the tangential component of the signed velocity almost vanishes indicating a tangential motion of the filaments in two opposing directions, while the perpendicular component of the signed velocity again indicates the net filament motion towards the membrane (blue region in Fig.~\ref{fig:density_time_evolution}h corresponding to the purple region in Fig.~\ref{fig:density_time_evolution}a). This motion leads to the accumulation of the filaments in the membrane vicinity. \par When the filaments accumulate in the membrane vicinity at the later stage $t \gtrsim 20$ , the polar order decreases and the nematic order appears. The filaments in the accumulation slide along the membrane as indicated by a finite value of their tangential speed. Note the similar motion was reported for self-propelled rods \cite{Abaurrea-Velasco2017Complex} although the propulsion mechanism is different from actomyosin. Since the filaments form nematic order, however, the tangential motion vanishes on average ($\langle v_{\parallel} \rangle \approx 0$) due to the existence of the counter-moving filaments. \par Finally, we estimate the width of the high nematic order region in the membrane vicinity. To this end, we measured the maximum nematic order parameter at each time after the system has reached a steady state (for $t\ge 40$). The time average of the maximum nematic order parameter is $\mathtt{n}_{\rm max} \approx 0.75$. We define the high nematic order region in the membrane vicinity as the region with $\mathtt{n} \ge \mathtt{n}_{\rm max} /2$. The boundary with $\mathtt{n} = \mathtt{n}_{\rm max} /2$ is displayed by the gray solid line in Fig.~\ref{fig:density_time_evolution}c. We estimate the width of the high nematic order region as the time-average of the boundary distance $d_{\rm memb}$. The obtained length scale is $d_{\rm memb} \approx 5.12$, which is plotted by the black dotted line in Fig.~\ref{fig:density_time_evolution}c. This width is approximately given by half the filament length $b_{\rm memb} = L_{\rm f}/2$, at which the filaments start to experience the depletion force. This agrees with the picture that the filaments are driven by the bulk polar order to accumulate in the membrane vicinity, and then exhibit a nematic order due to their interaction with the membrane.\section{Discussion} \label{sec:discussion}To summarize, we demonstrated that the competition between the depletion force and MWF leads to an accumulation of filaments in the membrane vicinity, which resembles the actomyosin cortex. The self-organized structure exerts pressure on the membrane, which converges to a constant value for large MWF because of nematic alignment of the filaments parallel to the membrane. Interestingly, for intermediate MWF, the pressure increases when the number of motors is decreased. We highlighted the universality of the phenomenon by showing the data collapse of the DDF peak position and the scaled pressure ($\mathcal{P} = P / N_{\rm f}$) as functions of the average MWF $\mathcal{F}^{\rm walk} = f^{\rm walk} N_{\rm m} / N_{\rm f}$. Qualitatively the same results are obtained for different membrane curvatures. Moreover, the results are quantitatively unaltered by the filament length, except for very short filaments for which the pressure profile changes while the structure formation is still observed. Our results provide a novel insight into the self-organization of cytoskeleton active matter into higher-order structures and its emergent mechanical function. \par Our model is relatively simple, such that our findings can in principle be tested against experiments on artificial cells \cite{Takiguchi2008Entrapping,Miyazaki2015Cell-sized}. However, to our knowledge, experiments on the self-organization of actomyosin cortex in artificial cells remain challenging \cite{Kurokawa2017DNA}, and oftentimes depletion agents are used instead to study the dynamics with the cortex. In comparison to the actomyosin cortex of real cells, our model omits many elements. For instance, actin nucleators and severing proteins are not included, which are thought to play a major role in the cortex formation \cite{Pontani2009Reconstitution}. Instead, we simplified the polymerization/de-polymerization processes with a stochastic turnover. As a result, the self-organized structures tend to fade away for very large turnover rates, leading to a uniform distribution of actin filaments. \par Recent studies reported that contraction occurs due to additional passive crosslinkers between filaments without force generating activity \cite{Hiraiwa2016Role, Kim2015Determinants,Belmonte2017A}. This effect also appeared in our simulation when crosslinkers were introduced, but was observed at positions both close to and far away from the membrane starting form a uniform distribution. This contraction occurs even without the membrane. Therefore, to reproduce a contractile actomyosin cortex in simulation, both motors and crosslinkers need to be introduced to a filamentous network that is enriched in the membrane vicinity due to membrane-associated nucleation and biased polymerization. To maintain the localization of the contractile actomyosin cortex in the membrane vicinity, additional linkers between the filaments and membranes are necessary. \par Finally, the extension of our system to three dimensions is important. In 3d, we expect defects to appear in the accumulation, which move dynamically. Such defects were previously found in the experiment using microtubules \cite{Sanchez2012Spontaneous}. Further, several studies using epithelial monolayers, another type of active system, focused on the dynamics of the defects, which causes cell exclusion and cell death due to the local stress~\cite{Saw2017Topological}. Therefore, it would be interesting to see if the defects in the accumulation of actomyosin also leads to the local change in the pressure. \begin{acknowledgments}This work was supported by the Japan Society for the Promotion of Science (JSPS) KAKENHI (19K14673 and 22K14017) grant, and RIKEN Special Doctoral Researcher (SPDR) Program. T.S. was supported by JSPS KAKENHI (JP19H00996) and JST CREST (JPMJCR1852), as well as by RIKEN Incentive Research Project and the core funding at RIKEN Center for Biosystems Dynamics Research. This study is initiated from the conversation among MT, ST, Yu-Chiun Wang, and Michiko Takeda. MT is grateful to Yu-Chiun Wang for stimulating discussion throughout this study and constructive comments on the manuscript from a biological viewpoint. MT acknowledges Matthew S. Turner and Sonja Tarama for careful reading and productive comments on the manuscript. \end{acknowledgments}M.T. designed the research with the help from T.S.; M.T. derived the model, developed the simulation code, and performed the analyses; M.T. prepared the manuscript; T.S. approved the final manuscript.\appendix\section{Equation of motion for motor head} \label{sec:EOM_motor}The position of the motor head $h$ ($=1,2$) of motor $m$ bound to a filament segment connecting $\bm{r}^{\rm f}_{i}$ and $\bm{r}^{\rm f}_{i+1}$ is given by \begin{align}\bm{r}^{\rm mot}_{m,h} = (1-\alpha_{m,h}) \bm{r}^{\rm f}_{i} +\alpha_{m,h} \bm{r}^{\rm f}_{i+1}, \label{eq:r^motor_h}\end{align}where $0\le \alpha_{m,h} \le 1$ measures the relative position along the filament segment. This allows us to take into consideration the situation where the motor head is bound to a position in between the filament particles. Then, the time derivative of Eq.~\eqref{eq:r^motor_h} consists of two terms: \begin{align}\frac{d \bm{r}^{\rm mot}_{m,h}}{dt} &= \bm{r}^{\rm f}_{i(i+1)} \frac{d \alpha_{m,h}}{dt} + \frac{d \bm{r}^{\rm f}_{f,\alpha_{m,h}}}{dt}.\label{eq:dr^motor_dalpha}\end{align}The first term represents the sliding motion of the motor head along the filament segment, whereas the second term \begin{align}\frac{d \bm{r}^{\rm f}_{\alpha_{m,h}}}{dt} = (1-\alpha_{m,h}) \frac{d \bm{r}^{\rm f}_{i}}{dt} +\alpha_{m,h} \frac{d \bm{r}^{\rm f}_{i+1}}{dt} \label{eq:r^fila_dalpha}\end{align}is the velocity of the filament at the motor head position. The latter corresponds to the transport of the motor head caused by the translation of the binding filament. Note that the left-hand side of Eq.~\eqref{eq:dr^motor_dalpha} is the velocity of the motor head measured in the Lab frame. \par Since we assume that the heads of the bound motors are always on the filaments, we only need to consider the time evolution of $\alpha_{m,h}$ to specify the position of the motor heads. The motor heads experience motor stretching and bending force as well as a motor walk force (MWF) $\bm{f}^{\rm mot,walk}_{m,h}$. We assign the former two to the particles of the filament segments to which the motor heads bind, to keep the motor heads on them. Then, the time-evolution equation of the motor head $h$ is given by\begin{align}\zeta \bm{r}^{\rm f}_{i(i+1)} \frac{d \alpha_{m,h}}{dt}&= \bm{f}^{\rm mot,walk}_{m,h}, \label{eq:dr^motor_h}\end{align}where $\zeta$ is the sliding friction coefficient between the motor head and filament. The molecular motor has an ability to walk actively along the filament by consuming chemical energy in the form of ATP. This effect is included in the model as the MWF \begin{align}\bm{f}^{\rm mot,walk}_{m,h} = \pm f^{\rm walk} \hat{\bm{r}}^{\rm f}_{i(i+1)}.\label{eq:f_motor_walk}\end{align}The sign specifies the direction of the motion; the plus (minus) sign corresponds to active walking towards the plus (minus) end of the filament. The direction of the motor walk is characteristic to the type of motor. In the case of the non-muscle myosin II motors that we consider in this study, the walking takes place towards the plus end of the actin filaments: $\bm{f}^{\rm mot,walk}_{m,h} = f^{\rm walk} \hat{\bm{r}}^{\rm f}_{i(i+1)}$. Here, we make a simplification on the process of motor head stepping forward along a filament, which is modelled by a constant walk force, since we are interested in the long time scale dynamics. This simplification allows us to neglect the transient unbind of the motor head while stepping forward. Note that, since the motor heads are moving along the filament, MWF acts in the direction parallel to the filament segments. \par The force from the motor head on the filament is given by\begin{align}\bm{f}^{\rm mot} ( \bm{r}^{\rm mot}_{m,h} )&= \zeta \bm{r}^{\rm f}_{i(i+1)} \frac{d \alpha_{m,h}}{dt}-\bm{f}^{\rm mot,walk}_{m,h}\notag\\&+ \bm{f}^{\rm mot,str}_{m,h}+ \bm{f}^{\rm mot,bend}_{m,h}.\label{eq:f^f_motor_h_full}\end{align}The first and second terms represent the counterpart of the sliding friction force between the motor head and filament and the MWF, respectively, so that the law of action and reaction is satisfied. The last two terms are the motor stretching and bending force acting on the motor head $h$, which is assigned on the filament particles. From Eq.~\eqref{eq:dr^motor_h}, however, the first two terms on the right-hand side of Eq.~\eqref{eq:f^f_motor_h_full} cancel and the remainder reads\begin{align}\bm{f}^{\rm mot} ( \bm{r}^{\rm mot}_{m,h} )&= \bm{f}^{\rm mot,str}_{m,h}+ \bm{f}^{\rm mot,bend}_{m,h}. \label{eq:f^f_motor_h}\end{align}Since the filament is modeled by discrete particles, this force should be split to the two edge particles ($\bm{r}^{\rm f}_{i}$ and $\bm{r}^{\rm f}_{i+1}$) depending on the geometric weight $\alpha_{m,h}$ as \begin{gather}\begin{array}{l}\bm{f}^{\rm f,mot}_{i} = (1-\alpha_{m,h}) \bm{f}^{\rm mot} ( \bm{r}^{\rm mot}_{m,h} ), \\ \bm{f}^{\rm f,mot}_{i+1} = \alpha_{m,h} \bm{f}^{\rm mot} ( \bm{r}^{\rm mot}_{m,h} ). \end{array}\label{eq:f^f_motor}\end{gather}See Fig.~\ref{fig:model}b. Note that $\alpha_{m,h}$ measures the position of the motor head bound to in between the filament particles. This assignment of the force ensures the conservation of the force and torque acting on the filament segment due to $\bm{f}^{\rm mot} ( \bm{r}^{\rm mot}_{m,h} )$. If there is more than one motor head on the filament segment $\bm{r}^{\rm f}_{i}$--$\bm{r}^{\rm f}_{i+1}$, then the right hand side of Eq.~\eqref{eq:f^f_motor} is summed up over them. As a result, the filament element $i$ experiences the force from the motors given by\begin{align}\bm{f}^{\rm f,mot}_{f,i} &= \sum_{(m,h) \in \mathcal{M}_{i-1}} \alpha_{m,h}\bm{f}^{\rm f,mot} ( \bm{r}^{\rm mot}_{m,h} )\notag\\&+\sum_{(m,h) \in \mathcal{M}_{i}} (1-\alpha_{m,h}) \bm{f}^{\rm f,mot} ( \bm{r}^{\rm mot}_{m,h} ),\label{eq:f^f_motor_sum}\end{align}where the first and second summations correspond to the contribution from the motor heads on the filament segments $\bm{r}^{\rm f}_{i-1}$--$\bm{r}^{\rm f}_{i}$ and $\bm{r}^{\rm f}_{i}$--$\bm{r}^{\rm f}_{i+1}$, respectively. Here, $\mathcal{M}_{i}$ represents the list of the motor heads bound to the filament segment $\bm{r}^{\rm f}_{i}$--$\bm{r}^{\rm f}_{i+1}$. \section{Parameter values} \label{sec:parameters}\begin{table*}[tb]\caption{The summary of the simulation parameters. The ones indicated in bold font are the input parameters of the simulation. }\centering\begin{tabular}{lllccc}&symbols && in silico& in vivo/in vitro&references\\ \hline\multicolumn{5}{l}{Cytosol} \\ \hspace{1em} \bf{thermal energy}&$k_B T$&& $4.142 \times 10^{-3}$ $\pN \, \mu\m$&at  $300$ $\textrm{K}$&\\ \hspace{1em} \bf{cytosolic viscosity}&$\eta_{\rm cyto}$&&$10^{-1}$ $\rm{Pa} \cdot \sec$&$10^{-3}$--$10^{-1}$ $\rm{Pa} \cdot \sec$&\cite{Valberg1987Magnetic} \\ \hline\multicolumn{5}{l}{Filament (actin)} \\ \hspace{1em} \bf{number of filaments}&$N_{\rm f}$&&200&--& \\ \hspace{1em} \bf{diameter}& $d^{\rm f}$&&0.01 $\mu\m$&0.009 $\mu\m$& \cite{Wen2011Polymer} \\ \hspace{1em} \bf{rest segment length}& $\ell^{\rm f}_0$&&0.2	 $\mu\m$&--& \\ \hspace{1em} \bf{number of elements}&$N^{\rm f,el}$&&11&--& \\ \hspace{2em} rest full length&$L^{\rm f}$&$=\ell^{\rm f}_0 (N^{\rm f,el}-1)$&2 $\mu\m$&$\lessapprox$ a few $\mu\m$& \cite{Mueller2017Load} \\ \hspace{1em} \bf{elastic modulus}&$\kappa^{\rm f,str}$&&10 $\pN$&& \\ \hspace{2em} Young's modulus&$E^{\rm f,str}$& $= \kappa^{\rm f,str} / \big( \pi (\frac{1}{2}{d^{\rm f}})^2 \big)$&$\sim 1.27 \times 10^5$ $\rm{Pa}$&$2 \times 10^{9}$ $\rm{Pa}$& \cite{Kojima1994Direct}\\ \hspace{1em} \bf{bending rigidity}&$k^{\rm f,bend}$&&0.1 $\pN \,\mu\m$&& \\ \hspace{2em} persistence length&$\ell^{\rm f,bend}$& $= k^{\rm f,bend} \ell^{\rm f}_0 / k_B T$&$\sim 24.1$ $\mu\m$&$\approx 20$ $\mu\m$&\cite{Wen2011Polymer} \\ \hspace{1em} \bf{turn over rate}&$\omega^{\rm f}_{\rm TO}$&&0.01 $\sec^{-1}$&& \\ \hline\multicolumn{5}{l}{Motors (non-muscle myosin II filaments)} \\ \hspace{1em} \bf{number of motors}&$N_{\rm m}$&&400&--&\\ \hspace{1em} \bf{diameter}&$d^{\rm mot}$&&0.01 $\mu\m$&$\approx$ 0.01--0.03 $\mu\m$	& \cite{Vasquez2016Drosophila} \\ \hspace{1em} \bf{rest length}&$\ell^{\rm mot}_0$&&0.2 $\mu\m$&$\approx$ 0.2--0.3 $\mu\m$&\cite{Vasquez2016Drosophila} \\ \hspace{1em} \bf{stretching modulus}&$\kappa^{\rm mot,str}$&&10	$\pN$&&\\ \hspace{2em} Young's modulus&$E^{\rm mot,str}$& $=\kappa^{\rm mot,str} / \big( \pi (\frac{1}{2} {d^{\rm mot}})^2 \big)$&$\sim 1.27 \times 10^5$ $\rm{Pa}$&\\ \hspace{1em} \bf{bending rigidity}&$k^{\rm mot,bend}$&& 0.1 $\pN \, \mu\m$&&\\ \hspace{2em} persistence length&$\ell^{\rm mot,bend}$&$= k^{\rm mot,bend} \ell^{\rm mot}_0 / 2 k_B T$& $\sim 24.1$ $\mu\m$&& \\ \hspace{1em} \bf{walk force}&$f^{\rm mot,w}$&&0.05 $\pN$&$\lessapprox $ a few $\pN$ (stall force)&\cite{Chaen1995The}\\ \hspace{1em} \bf{sliding friction}&$\zeta$&&0.1 $\unitfric$&$\lessapprox 0.1 \,\unitfric$&\cite{Tawada1991A} \\ \hspace{1em} \bf{turn over rate}&$\omega^{\rm mot}_{\rm TO}$	&&0.01 $\sec^{-1}$&&\\ \hline\multicolumn{5}{l}{Repulsive interaction between filaments and membrane}\\ \hspace{1em} \bf{modulus}&$\epsilon^{\rm fila:memb}$&&10 $\pN \, \mu\m$&--& \\ \hspace{1em} \bf{length}&$\sigma^{\rm fila:memb}$&&0.05 $\mu\m$&--&\\ \hspace{1em} \bf{cutoff length}&$r^{\rm fila:memb}_{\rm cutoff}$&&$4.2$ $\sigma^{\rm fila:memb} = 0.21$ $\mu\m$&--&\end{tabular}\label{table:parameters}\end{table*}The simulation parameters for the filaments and motors are set as summarized in Table~\ref{table:parameters}, unless otherwise stated. Some of the parameters are compared with existing experimental measurements. The bending rigidity of the motors is set to the value same as that of the filaments, since, to our knowledge, it has not been measured for myosin minifilaments yet. The parameters of the confining membrane and system size are set as follows. For the circular membrane with negative curvature, the radius is set as $R = 20$, which leads to the curvature $\kappa = -0.05$. For the plane membrane with zero curvature $\kappa=0$, the distance between membranes is $L_x = 30$ and the length of the membrane $L_y = 41.75$ with periodic boundaries. For the circular membrane with positive curvature, in which case the filaments are present around the membrane, the radius is $R = 20$, and the width and height of the simulation box is set $L_x = L_y = 70.9$ with periodic boundaries. This leads to the curvature $\kappa = 0.05$. \section{Depletion force} \label{sec:depletion}
\begin{figure*}[tb]\centering\includegraphics{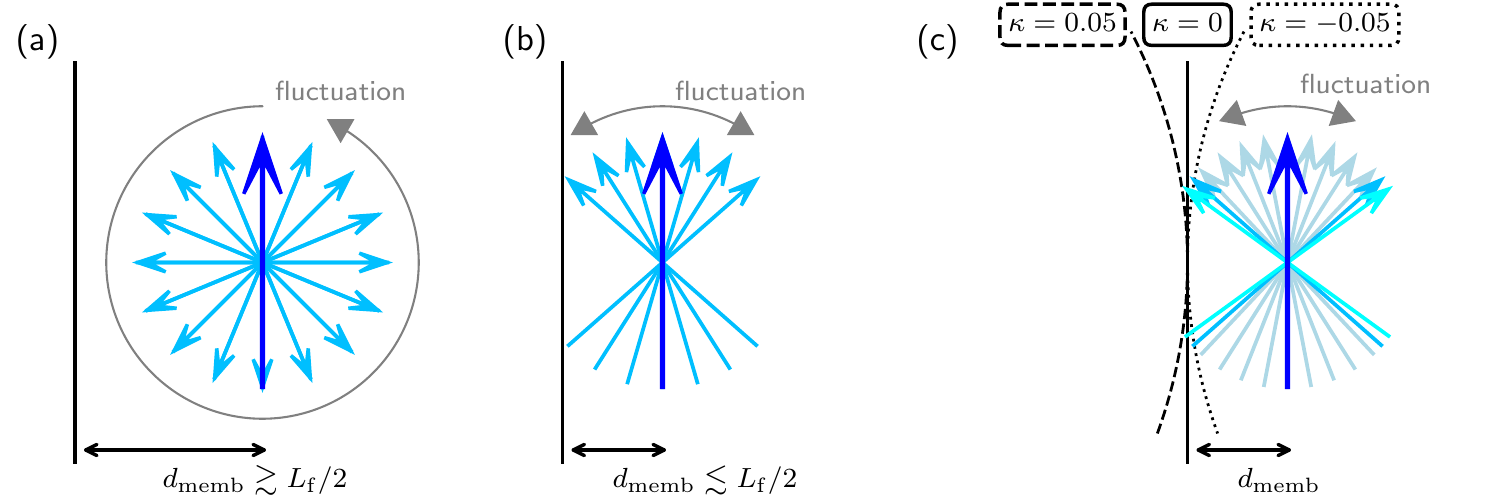}\caption{Schematics of the rotation of a filament (blue) due to thermal fluctuation depending on its relative position to the membrane (black). Filament placed (a) far from and (b) close to the membrane. A filament placed far from membrane can rotate freely, while the rotational degree of freedom of a filament placed close to the membrane is restricted, which gives rise to a depletion force. The rotation of a filament placed parallel to the membrane is indicated by the light blue arrows. The threshold distance at which the filament starts to experience depletion force is given by half of the filament length $L_{\rm f}/2$. (c) Filament has more rotational degrees of freedom as the membrane curvature increases. The thin ash blue arrows indicate the rotation of the filaments around the membrane with $\kappa=-0.05$, and the light blue and cyan arrows represent the additional rotational degrees of freedom around the membrane with $\kappa=0$ and $\kappa=0.05$, respectively. The dotted, solid, and dashed black lines represent the membrane with curvature $\kappa = -0.05$, 0, and 0.05, respectively. }\label{fig:depletion}\end{figure*}

To provide an intuitive idea of the depletion force, we show schematic sketches depicting a filament placed far from and close to the membrane in Fig.~\ref{fig:depletion}. A filament placed far from the membrane can rotate freely due to thermal noise (Fig.~\ref{fig:depletion}a), whereas the rotational degree of freedom of a filament placed close to the membrane is restricted (Fig.~\ref{fig:depletion}b). This entropic penalty tries to keep the filament away from the membrane, giving rise to a depletion force. The threshold distance at which the filament starts to experience the depletion force is approximately given by half of the filament length $d_{\rm memb} = L_{\rm f} /2$. The cutoff distance of the repulsive interaction between the membrane and filament particles lies within the threshold distance in this study. \par In Fig.~\ref{fig:depletion}c, we show the effect of the membrane curvature on the depletion force. As the membrane curvature decreases, the filament rotation is more restricted. Therefore, the filaments experience a stronger depletion force for decreasing the membrane curvature. As a result, in the membrane vicinity, filaments originally placed parallel to the membrane at the same distance are more likely to point away from the membrane due to fluctuation as the membrane curvature increases. 
\end{document}